\documentclass{article}

\usepackage[final]{aloe_2023_neurips}

\usepackage[utf8]{inputenc} 
\usepackage[T1]{fontenc}    
\usepackage{hyperref}       
\usepackage{url}            
\usepackage{booktabs}       
\usepackage{amsfonts}       
\usepackage{nicefrac}       
\usepackage{microtype}      
\usepackage{xcolor}         
\usepackage{amsmath}        
\usepackage{graphicx}       
\usepackage{subcaption}     
\usepackage{enumitem}       
\usepackage{hyperref}
\usepackage{gensymb}      
\usepackage{wrapfig}
\usepackage{graphicx}

\title{From Centralized to Self-Supervised: Pursuing Realistic Multi-Agent Reinforcement Learning}

\author{
 Violet Xiang\thanks{Corresponding author: ziyxiang@stanford.edu} \quad
  Logan Cross \quad
  Jan-Philipp Fränken \quad
  Nick Haber \\
  Stanford University
}
\begin{document}

\maketitle

\begin{abstract}
    In real-world environments, autonomous agents rely on their egocentric observations. 
    They must learn adaptive strategies to interact with others who possess mixed motivations, discernible only through visible cues.
    Several Multi-Agent Reinforcement Learning (MARL) methods adopt centralized approaches that involve either centralized training or reward-sharing, often violating the realistic ways in which living organisms, like animals or humans, process information and interact.
    MARL strategies deploying decentralized training with intrinsic motivation offer a self-supervised approach, enable agents to develop flexible social strategies through the interaction of autonomous agents. 
    However, by contrasting the self-supervised and centralized methods, we reveal that populations trained with reward-sharing methods surpass those using self-supervised methods in a mixed-motive environment.
    We link this superiority to specialized role emergence and an agent's expertise in its role. 
    Interestingly, this gap shrinks in pure-motive settings, emphasizing the need for evaluations in more complex, realistic environments (mixed-motive).
    Our preliminary results suggest a gap in population performance that can be closed by improving self-supervised methods and thereby pushing MARL closer to real-world readiness.
\end{abstract}

\section{Introduction}
\label{intro}
    In real-world environments, agents' ability to successfully operate depends on how they navigate and interpret interactions with other agents and entities they encounter.
    The dynamics of these interactions are subject to change over time and vary among individuals, as different agents may have divergent or mixed motives influencing their behavior.
    For instance, vendors in a farmer's market both compete for customers and cooperate with nearby stalls for shared resources. 
    Their interactions shift based on immediate needs, like maximizing sales or maintaining good relations, showcasing the fluid nature of mixed motives in real-world settings.
    These interactions are often driven by underlying motivations, social norms, and evolving relationships.
    
\begin{figure}[h]
        \centering
         \includegraphics[width=.95\textwidth]{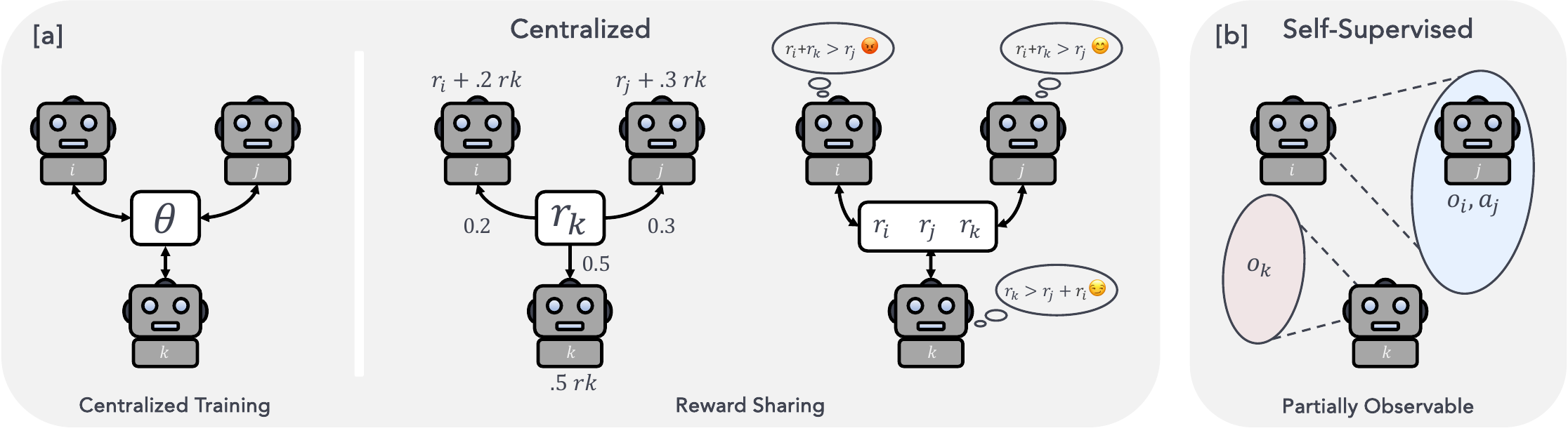}
        \caption{Illustration of different MARL approaches. 
        \textbf{Centralized Training}: Agents share parameters, either policy or critic.
        \textbf{Reward sharing}: Method 1: Agents distribute their rewards amongst themselves to influence behavior.
        Method 2: Agents are aware of others' rewards and adjust their actions accordingly. An added term to their reward, such as social value orientation, serves as an intrinsic motivation, determining if an agent values the success of others over its own.
        \textbf{Self-supervised}: These approaches operate with minimal assumptions, as agents must infer hidden information about other agents solely from their partially observable egocentric observations.}
        \label{fig:1}
    \end{figure}

    For agents to be efficient, they must not only witness but also comprehend and adjust to these intricate dynamics.    
    Existing multi-agent system solutions predominantly employ centralized strategies, as illustrated in the left panel of Fig \ref{fig:1}. 
    These can involve a centralized training through sharing parameters \citep{lowe2017multi, yu2022surprising} or sharing information that individual agents would not typically access without explicit communication or insights into other agents' non-observable information \citep{hughes2018inequity, mckee2020social, mckee2021multi, baker2020emergent}.   
    Centralized methods leverage shared information among a cohort of agents to promote coordination.
    However, the real world places agents together with AI entities beyond their training cohort or even humans, underscoring the need for self-supervised learning, akin to biological systems.
    In such settings, agents have limited information, relying on observable cues to infer states, objectives, and rewards of others. 
    To make optimal decisions, they must discern the intentions of others and group dynamics based solely on this observable information.
    
    \paragraph{} This observational challenge accentuates the importance of adaptive, self-supervised strategies in realistic multi-agent settings, illustrated in Fig \ref{fig:1}b. 
    A cornerstone of self-supervised approaches is intrinsic motivation, a force that drives actions without external rewards, especially crucial in sparse-reward environments \citep{pathak2017curiosity}.
    Methods using curiosity as intrinsic motivation promote adaptability across diverse tasks and environments by allowing an agent to seek more information about its environment by exploring unfamiliar states and improve its model of the environment \citep{pathak2017curiosity, haber2018learning, burda2018exploration, pathak2019self, kim2020active, kauvar2023curious}. 
    While these epistemic intrinsic motivations effectively drive agents to improve their model of the environment, they might not always offer definitive guidance on how agents should behave in relation to others, i.e., helping or undermining others.
    \citep{ndousse2021emergent} show that integrating an model-based auxiliary loss, allowing agents to predict upcoming states from observations, promotes social learning in certain environment even without intrinsic motivation.
    Moreover, model-based intrinsic motivations that emphasize interactions with other agents provide insights into how agents can better model peer behaviors.
    Such motivations involve predicting other agents' actions or states, enhancing an agent's ability to align its policy with the behaviors of surrounding entities.
    These motivations, whether for exploration, or social influence, guide decisions and interactions, pushing individuals to cooperate, compete, or even conform in search of social equilibrium \citep{jaques2019social, ma2022elign}.

    While studies have explored centralized, epistemic, and social intrinsic motivations, there is a clear need for a head-to-head comparison within a consistent environment and agent setup.
    Conducting such an evaluation in a uniform setting isolates the impact of each motivation type, ensuring performance or behavioral differences arise from the motivations themselves and not extraneous factors. 
    Our research bridges this gap. 
    We provide a direct comparison in two distinct environments: one with mixed-motive dynamics and another with pure-motive dynamics, all within a low sample complexity setting.
    Evaluating in this particular regime ensures readiness for the real world which is characterized by its unpredictable nature making rapid adaptability a paramount feature.
    Agents capable of learning efficiently from minimal samples can swiftly adapt to emerging challenges without the requirement of extensive retraining.
    Similar to biological organisms, agents in real-world scenarios would benefit immensely from continual and lifelong learning, where the ability to learn and adapt consistently over extended periods without vast new data influxes is invaluable.
    This sheds light on performance differences and population composition, highlighting the promise of intrinsic motivation in autonomous agent design.
    Our contributions are twofold:
    \begin{enumerate}
        \item We identify the disparities among centralized and self-supervised MARL models, highlighting areas for future research.
        \item We discern the factors behind the population outcome of various models, identifying a critical need for specialized roles and individual agents' competence in fulfilling those roles.        
    \end{enumerate}

\section{Background}
\label{related}
\paragraph{Centralized MARL} Centralized Multi-Agent Reinforcement Learning (MARL) often trains agents in a collective, targeting optimal performance.
Method by \citep{lowe2017multi}, individualizes agent policies but centralizes training data with a shared critic.
Method by\citep{yu2022surprising} shares a policy network across agents, optimizing actions for aggregate returns. 
Some strategies permit decentralized learning but demand unrealistic data access.
For example, \citep{mckee2021multi} uses a reputation-based motivation, specific to the Clean Up environment, thereby constraining its versatility. \citep{mckee2020social} applies 'social value orientation' (SVO) to allocate altruistic or selfish tendencies. 
Here, each agent's SVO is set in a diverse population, where the SVO reflects a balance between maximizing an agent's own reward and improving rewards for fellow agents. 
While this fosters cooperation, it requires agents to know peers' rewards, a limitation somewhat sidestepped by \citep{li2023learning} through allowing agents to distribute their own reward, but at the cost of explicit inter-agent communication.
Our focus remains on agents understanding peers strictly via direct observations.

\paragraph{Intrinsic motivation in single and multi-agent reinforcement learning}

Intrinsic motivation guides agents to effectively navigate open-ended environments \citep{oudeyer2007intrinsic, schmidhuber1991possibility}.
One approach is to set and achieve self-generated sub-goals coinciding with environment tasks \citep{lair2019language, campero2020learning, forestier2022intrinsically, colas2022autotelic}.
Another approach, empowerment \citep{klyubin2005empowerment}, increases an agent's control over the environment by maximizing the mutual information between state and action \citep{choi2021variational}.
Curiosity is a different form of intrinsic motivation promoting information-seeking behavior in an agent.
Building on curiosity, \citep{pathak2017curiosity} rewards exploration based on forward prediction losses. 
\citep{pathak2019self} targets states with high uncertainty, and \citep{kim2020active} refines agent models with a curiosity signal. 
Extending to multi-agent settings, we analyze with ICM motivation.
Besides these epistemic motivations, there exist motivations tailored to the social intricacies of multi-agent interactions.
\citep{jaques2019social} introduces social influence as a motivation, prompting agents to increase their influence on others' actions. 
An agent's social influence rewarded is defined by the difference between other agents taking certain actions given its action and without its action.
\citep{ma2022elign} promotes spontaneous coordination by motivating them to behave in ways that are most expected by other agents using only local observations and actions.
Though these methods vary, direct comparisons in a unified setting are lacking. Our research offers this comparison, highlighting the strengths and weaknesses of each motivation.

\section{Experimental Settings}
\subsection{Environment}
\begin{wrapfigure}{r}{0.5\textwidth}
    \centering
    \includegraphics[width=.95\linewidth]{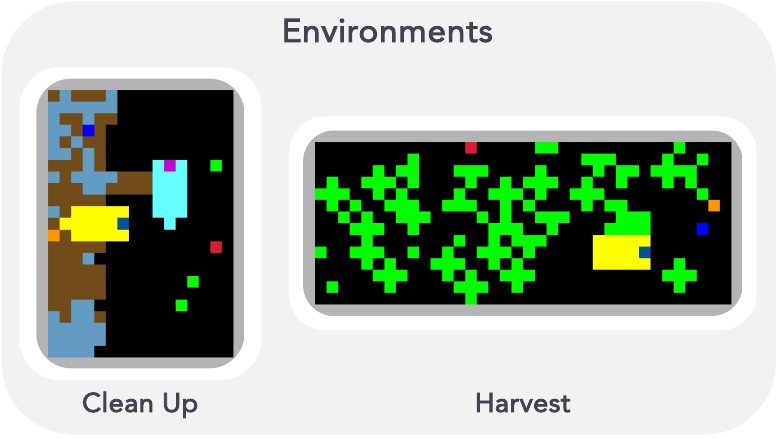}
    \caption{Clean Up (left) map size is 25x18. Harvest (right) map size is 38x16.} 
    \label{fig:2}
\end{wrapfigure}

To evaluate model performance across varied social dynamics, we selected both mixed-motive (Cleanup) and pure-motive (Harvest) environments as proposed by \citep{hughes2018inequity}.
Examples of these environments are shown in Fig \ref{fig:2} \footnote{We adapted environment implementation from \href{https://github.com/eugenevinitsky/sequential_social_dilemma_games}{this repository}}. 
Unlike the standard matrix games in game theory, these environments utilize spatially and temporally extended games with embodied avatars.
In Clean Up (Fig \ref{fig:2} left), agents are 
rewarded for consuming apples in an environment with a polluted river.
Initially, apples are not available on the map until agents start to clean the river. 
Apple growth is tied to river cleanliness, with growth rates dropping as pollution increases. 
Additionally, the apple growth area is distant from the river, requiring agents to balance between cleaning and apple consumption.
This poses an challenging dynamic as when they are multiple agents in the environment, agents will have conflicting motivations (consuming apples vs cleaning the river).
In Harvest (Fig \ref{fig:2} right), agents are rewarded for apple consumption in an environment populated with apples. Apple growth relies on the proximity of other apples, and over-harvesting leads to non-regeneration. This environment thereby symbolizes the tragedy of the commons.
All models receive inputs in the form of 15x15 square observations centered around the agents. 
The policy network for every agent consists a convolutional neural network (CNN) followed by a gated recurrent unit (GRU) network. The ICM, ICM-reward, and social influence agents incorporate distinct world model components for their respective predictions.
For ICM and ICM-reward agents, a separate CNN encoder, coupled with a GRU, predicts environmental dynamics.
For social influence agents we follow the setup in the original paper \citep{jaques2019social} by sharing the same CNN encoder as the policy and adding a separate head with two MLP layers and a GRU as the MOA model.
Differing from only giving one or a few agents with the ability to influence other agents, we allow all agents in a population to be rewarded for influencing other agents visible to them. 
Each model is trained with episode length of 2000 steps (Clean Up) or 1000 steps (Harvest), and Bayesian hyper-parameter search is performed for each model to ensure the best performance. The results in the section \ref{res} are from the best hyperparameters for each model.

\subsection{Evaluated Models}
A promising model should allow agents to discern optimal strategies depending on the environment structure, constraints, and emergent social dynamics.
To capture these constraints in a realistic setting, we formulate a Markov game for \(K\) agents as \(<S, A, R, P>\). Here:
\begin{itemize}
    \item \(S\) represents the environment state.
    \item \(A\) denotes the joint action set of all agents.
    \item \(R\) is the reward function, defined as \(R(S, A) = \{r_1, ..., r_K\}\).
    \item \(P\) outlines the state transition dynamics, given by \(P(S, A) = S'\).
\end{itemize}
At each timestep \(t\), an agent \(i\) accesses its state \(s_i^t\) via observation \(o_i^t\), takes an action \(a_i^t\), and receives reward \(r_i^t\). 
Additionally, the agent observes actions \(a_j^t\) of other agents within its neighborhood, defined by \(N(i)\).
We evaluated these models:
\begin{itemize}
    \item \textbf{Independent PPO (IPPO) agents} each have their own policy and value network trained through the proximal policy optimization (PPO) algorithm \citep{schulman2017proximal}, targeting the maximization of individual extrinsic rewards from the environment. They serve as solvers for other models with intrinsic rewards. 
    \item \textbf{MAPPO agents} use the same solver as the IPPO agents while sharing a single value and policy network \citep{yu2022surprising}. The policy network is trained with trajectories collected from all agents to provide actions based on individual agents' local observations, and the value network is trained using the global state.
    \item \textbf{ICM agents} each contains a policy parameterized by ${\theta}$ and a world model parameterized by ${\phi}$ computing forward and inverse dynamics. The world model $f_{\phi}$ is trained by simultaneously minimizing a forward prediction loss
        \begin{align}
            L_{forward} = \| \mathbf{f_{\phi}(o_i^{t}, a_i^{t})} - \mathbf{o_i^{t+1}} \|
        \end{align}
        and an inverse dynamics through separate heads
        \begin{align}
            L_{inverse} = \mathbf{cross\_entropy}(f_{\phi}(o_i^{t}, o_i^{t+1}),  a_i^{t})
        \end{align}
        Each agent learns its policy based on the following reward formulation using the forward dynamics loss as an intrinsic reward as it is relevant for policy exploration.
        \begin{align}
            r_i = r_{i_{ext}} + \alpha r_{i_{int}} \text{  where  } r_{i_{int}}=L_{forward}
        \end{align}
    \item \textbf{ICM-reward agents} are trained similarly as the ICM agents mentioned above with an extra loss for the dynamic model to predict the reward
        \begin{align}
            L_{reward} = \| \mathbf{f_{\phi}(o_i^{t}, a_i^{t})} - \mathbf{r_i^{t+1}} \|
        \end{align}
    while the policy is optimized with this reward
        \begin{align}
            r_i = r_{i_{ext}} + \alpha r_{i_{int}} \text{  where  } r_{i_{int}}=L_{reward}
        \end{align}
    \item \textbf{Social influence agents}: These agents aim to optimize their influence on other agents' behaviors. 
    Each agent's Model of Agents (MOA) predicts actions of peers based on current observations and prior actions:
        \begin{align}
            L_{moa}(i) = \mathbf{cross\_entropy(f(o_i^{t}, A^{t-1}), A_{j \neq i}^{t}}) \texttt{ where } A=\{ a_j \texttt{ for } j \in N(i) \}
        \end{align}
    The MOA is used to compute the influence reward by computing the discrepancy between the marginal policy of other agents and the conditional policy of other agents given the self agent's action. 
    Due to the constraint of strict partial observability, agents use their observation to compute their influence over all other agents that are within their visibility radius.
    This incentivizes agents to increase their action's causal influence over other agents' actions which could lead to coordinated behaviors even without specific population-level control.
    This is computing the intrinsic reward:
        \begin{align}
            c_{i}^{t} = \sum_{j \neq i}^{K} (D_{KL}[\mathbf{p(a^{t}_{j} | a^{t}_{i}, o^{t}_{i}) \| p(a^{t}_{j}|o^{t}_{i})}])
        \end{align}
    Individual agents' policy is then optimized using this reward formulation
        \begin{align}
            r_i = r_{ext} + \alpha r_{int} \text{ where } r_{int} = c_i
        \end{align}
    \item \textbf{SVO agents} employ a centralized method, emphasizing reward sharing, which requires agent awareness of others' rewards at every step. 
    SVO modulates reward preferences through sampling SVO values from a preset distribution, and each agent's SVO determines their inclination to optimize others' rewards versus their own.
    The higher its SVO, the more altruistic an agent is. 
        \begin{align}
            SVO^{t}_{i} = arctan(\frac{\frac{1}{N-1} \sum_{j \neq i}^{K} r^{t}_{j}}{r^{t}_{i}} )
        \end{align}
    
    Given the pre-specified SVO value ($SVO_i^{'}$) sampled from the distribution $N(\mu, \sigma)$ for each agent $i$, its policy is optimized using such reward
        \begin{align}            
            r^{t}_{i} = r_{ext} - \alpha |SVO^{'}_i - \mathbf{clip}(SVO^{t}_{i}) | 
        \end{align}
    where the clip function bounds \(SVO^{t}_{i}\) within [0, \(\pi/2\)]. 
    We evaluate both homogeneous populations (\textbf{SVO-HO}) with consistent SVO values, and heterogeneous ones (\textbf{SVO-HE}) with diverse SVOs sampled from a distribution.
\end{itemize}

\section{Results}
\label{res}
\begin{figure}[h]
    \centering
     \includegraphics[width=.95\textwidth]{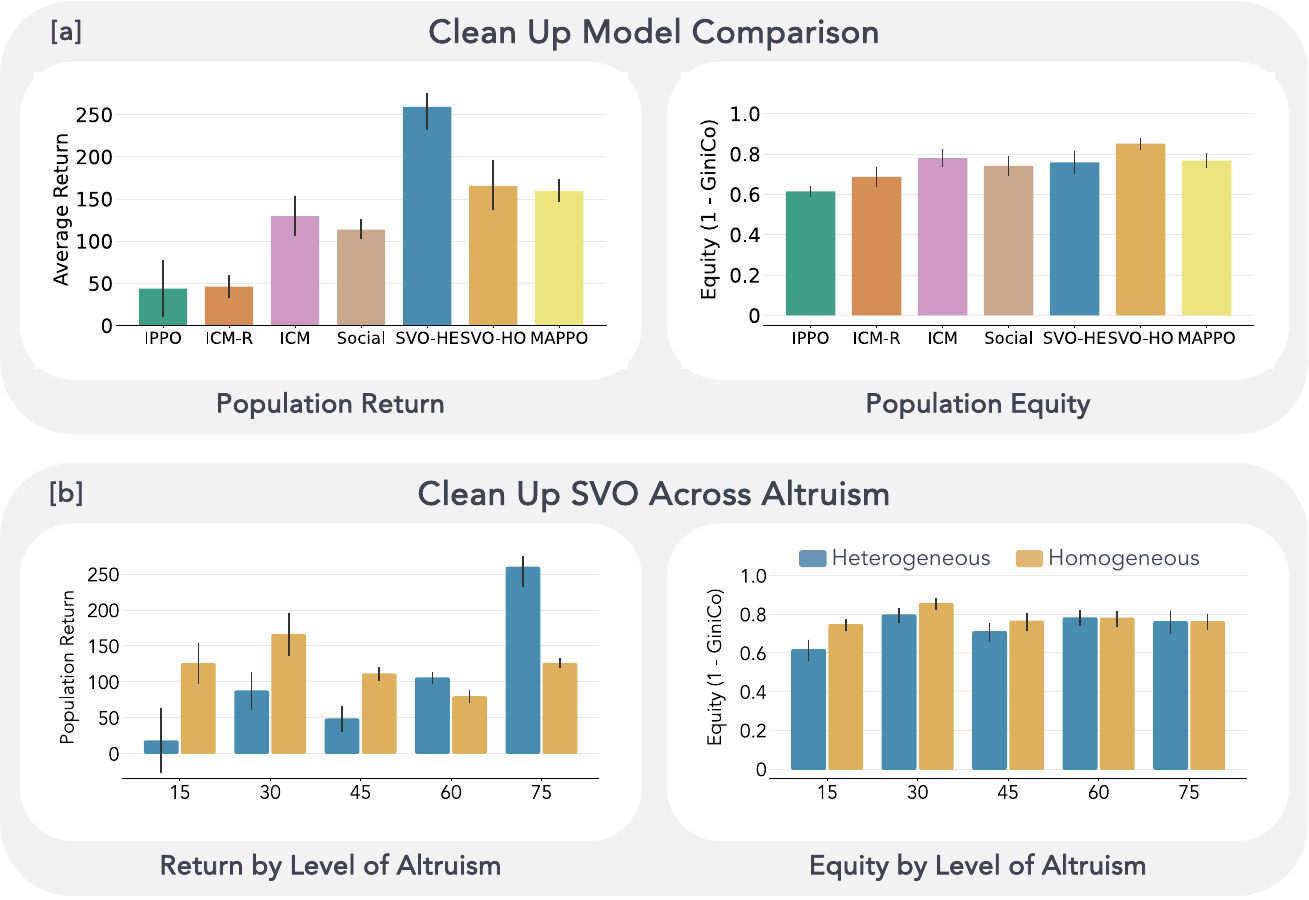}
     \caption{Clean Up Results. (a) Population return and equity score computed as 1 - Gini Coefficient, (b) Return and equity by SVO values. The error bars are standard error.}
    \label{fig:3}
\end{figure}

We trained populations of five agents for each model and evaluated them across five episodes.
The agents are trained up to one million steps but the best performing epoch, defined as every 5000 environment steps, was selected for each model (specific environment number are available in supplementary material \ref{supp}).
Average population (Fig \ref{fig:2}a) and equity is depicted in Fig \ref{fig:2}a right panel. To evaluate how equitable the return is across the 5 agents, the Gini coefficient is computed as $\frac{\sum_{i=1}^k \sum_{k=1}^k |r_i - r_j|}{2m\sum_{i=1}^k}$ as in \citep{leibo2021scalable} with a minor difference in handling negative reward values.
Instead of only using positive reward, we shift the minimum negative reward to zero plus a small $\delta=0.0001$ to ensure that negative returns also contribute to equity computation.

\paragraph{Clean Up} Centralized methods (SVO-HE, SVO-HO, MAPPO) are better than the all other self-supervised methods as shown in Fig \ref{fig:3}a.
SVO-HE outperforms other centralized methods by a large margin.
As expected, baseline IPPO agents exhibited the weakest performance. 
Agents with intrinsic motivations showed a marginal improvement over the IPPO baseline.
Interestingly, models using explicit social intrinsic motivation did not surpass those using non-social intrinsic motivations as ICM performance is very comparable to Social Influence model as show in Fig \ref{fig:3}a.
While achieving high rewards is a priority, it is vital to ensure minimal disparities among agents' payouts in the environment.
From Fig \ref{fig:3}a, it is evident that the SVO-HO model achieved the highest Gini coefficient scores and thus the most equitable reward distribution.
However, its overall performance, considering the population returns, was outperformed by the SVO-HE model which has more altruistic tendency in the optimal SVO hyperparameter and a diverse population.

Two variations of the SVO models are depicted in Fig \ref{fig:3}a heterogeneous population, where the SVO values for five agents are drawn from an Gaussian distribution with parameters $\mu=75\degree, \sigma=11.9\degree$, which is very altrusitic, and a homogeneous population featuring five agents with $30\degree$ SVO values. These values were selected by a hyperparameter search.
The heterogeneous SVO agents outperformed their homogeneous counterparts, aligning with previous findings \citep{mckee2020social} that suggest diverse SVO populations excel in mixed-motive settings compared to purely cooperative groups.
To delve deeper into the influence of population SVO configurations, we analyzed their impact on population returns and equity (Fig \ref{fig:3}b). 
Interestingly, altruism levels do not directly correlate with returns or equity. 
Analyzing the population return data, we observed that when agents converge to a high-performing population, they excel; otherwise, performance tends to be suboptimal, suggesting that hyperparameter selection is crucial.

\begin{figure}[h!]
    \centering    
     \includegraphics[width=.95\textwidth]{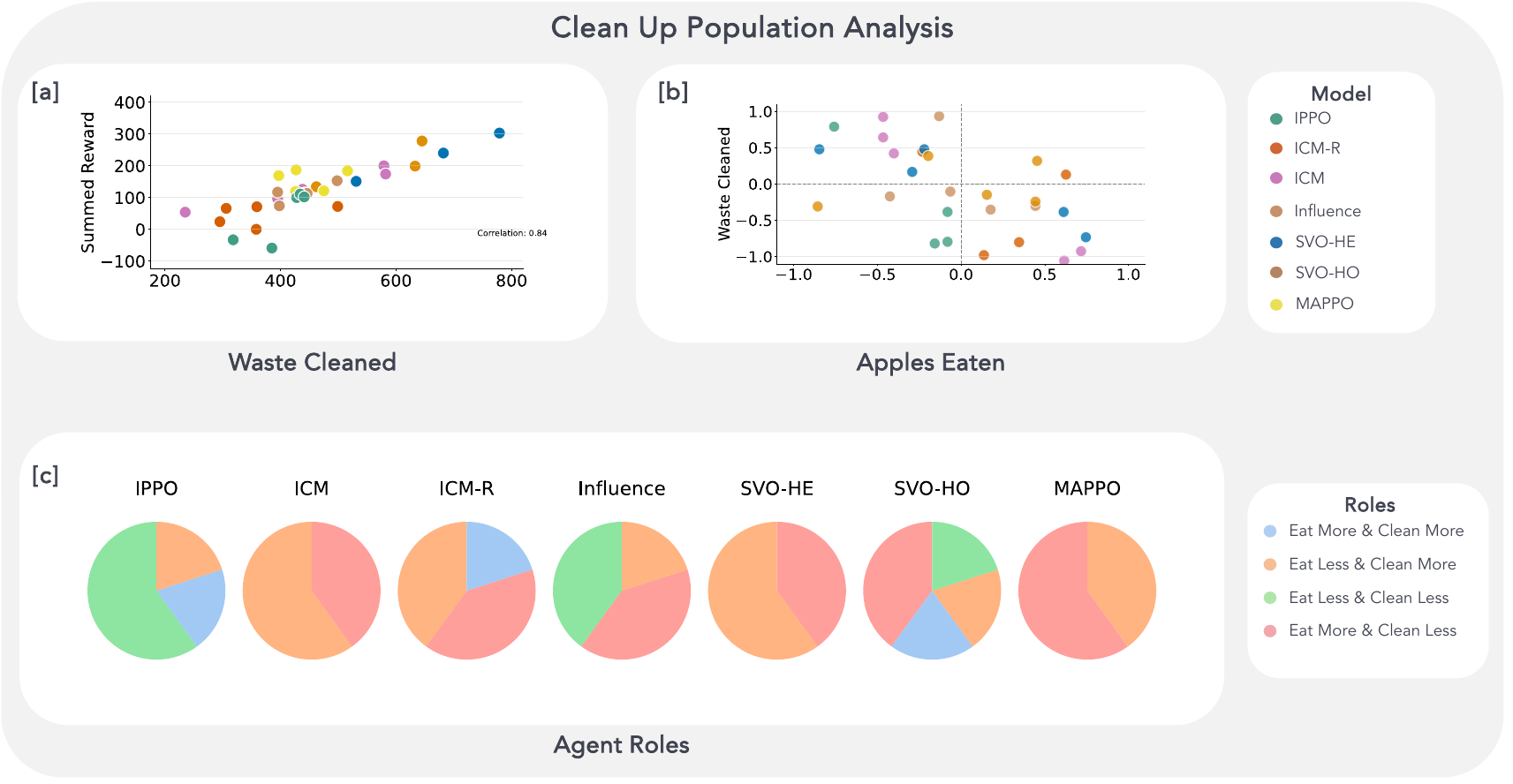}
    \caption{\textbf{(a)} Correlation ($r=0.88$) between the number of waste cleaned (x-axis) and the population reward (y-axis). \textbf{(b)} z-scored apple eaten (x-axis) vs waste cleaned (y-axis), this divides agents into four types by quadrant: top-right Eat More \& Clean More, top-left Eat Less \& Clean Less, bottom-left Eat Less \& Clean Less and Eat More \& Clean Less. \textbf{(c)} Roles in different population. Each role is defined by which one of the four quadrant an agent is fallen into on (b).}
    \label{fig:4}
\end{figure}

To discern the factors impacting the performance of different models, we analyze the relationship of cooperative behavior with population return.
Fig \ref{fig:4}a reveals a direct correlation ($r=0.88$) between waste cleaned and rewards, suggesting that it is advantageous for everyone when a group collectively cleans waste. 
We categorized agents based on their apple consumption and waste cleaning relative to the population average (Fig \ref{fig:4}b\&c). 
Although one IPPO agent outperformed its peers, the IPPO population lagged behind SVO-HE, ICM, and Social Influence due to three underperforming agents. 
Interestingly, ICM and SVO-HE both had three agents prioritizing cleaning and two focusing on eating. 
However, SVO-HE agents were more proficient and specialized, as evident in Fig \ref{fig:4}a,c, as agents were split between eating and cleaning roles. 
This underscores that a population's effectiveness hinges on agents specializing in roles and excelling in them.

\begin{figure}[h]
    \centering
     \includegraphics[width=.95\textwidth]{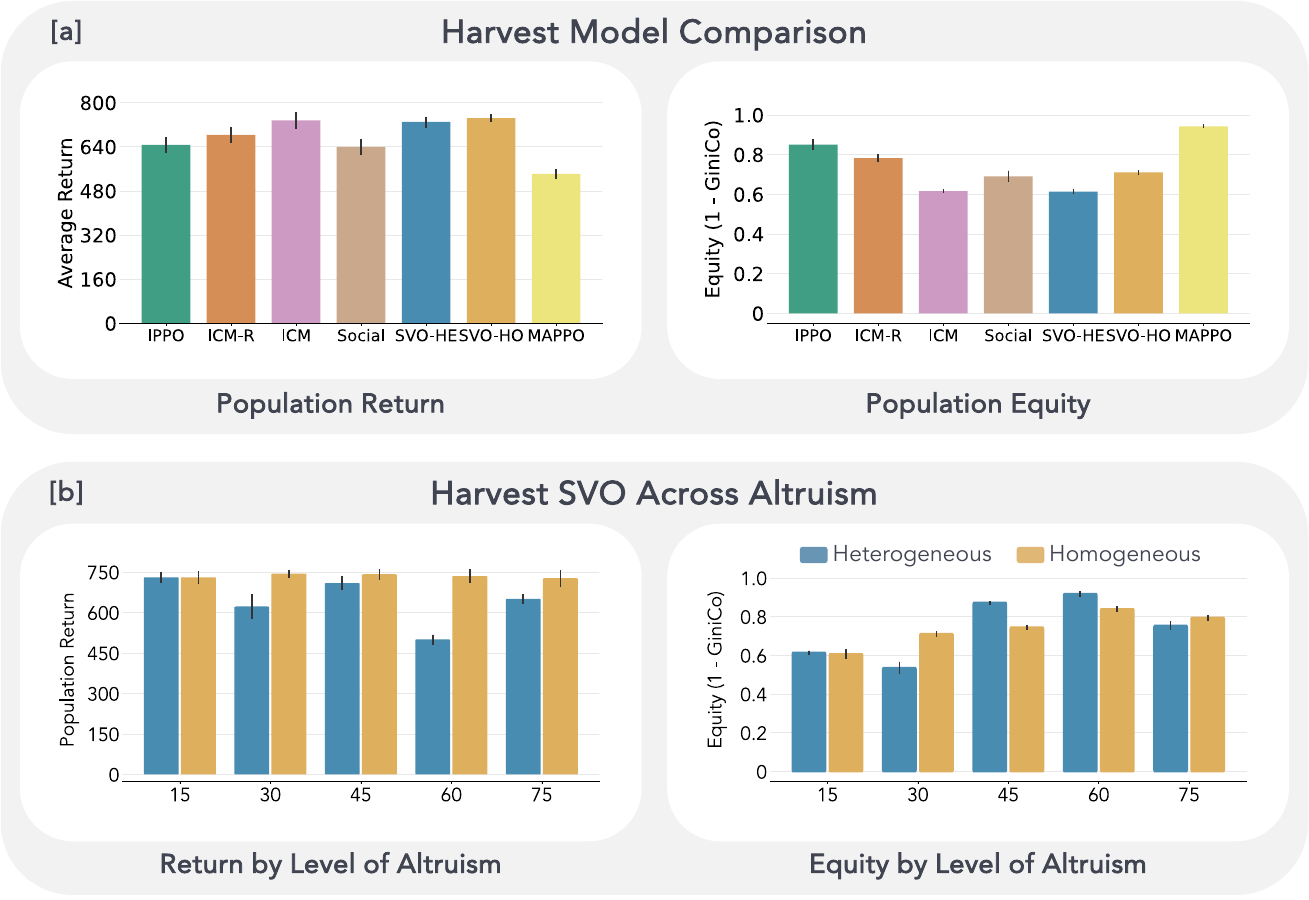}
     \caption{Harvest Results. (a) Population return and equity score computed as 1 - Gini Coefficient, (b) Return and equity by SVO values. The error bars are standard error.}
    \label{fig:5}
\end{figure}

\paragraph{Harvest} We also evaluated all models in Harvest, a pure motive environment distinct from Clean Up where every agent has an incentive to maintain sustainable eating.
In Harvest, dense apples are readily available without any preliminary tasks, such as cleaning the river in Clean Up, making it easier to achieve higher population returns (Fig \ref{fig:2}).
Fig \ref{fig:5} shows SVO-HO being the best model, however, there is no substantial difference between SVO models and others while MAPPO agents perform significantly worse than other model.
This highlights the need for evaluation in environments with mixed-motive social dynamics.
Notably, the optimal SVO parameter in a heterogeneous population differ between Harvest ($\mu=15$) and Clean Up ($\mu=75$), highlighting the challenge of pre-specifying a population across environments.
An ideal model should excel in both settings with the same hyperparameter, adapting its policy through agent interactions. 
Our study indicates the need for developing such versatile models.
\label{res}

\section{Discussion}
We assess various multi-agent models in two environments, each possessing distinct social dynamics.
Centralized approaches surpassed self-supervised approaches in the mixed-motive setting Clean Up but depends on pre-set population parameters.
When transitioning to a pure motive environment, these optimal parameters shift. 
This suggests that population reconfiguration is needed if environmental dynamics change, highlighting its limitation in allowing agents to flexibly achieve optimal performance. Additionally, this method relies on reward sharing mechanisms that may not be possible in every environment. 
In Clean Up, a notable gap exists between agents driven by self-supervised intrinsic motivations and those trained with centralized methods. 
Other social intrinsic motivations might bridge this gap.
The analysis of different population composition in Clean Up further suggests the effectiveness of a multi-agent population is strongly influenced by agents' role specialization and their proficiency in those roles.
The similar performance of methods in Harvest indicates the importance of evaluation in mixed-motive environments.
Due to resource constraints these models are trained under a low-sample regime. A major current limitation in MARL is the large number of samples that is required for populations to converge and reach equilibrium seen in cooperative dynamics. 
This provides additional motivation for investing in methods that can be flexible across environments without the required population pre-configuration and extensive hyper-parameter searches. 
Additionally methods need to be adaptable whenever there is a dynamics change in the environment. 
Despite these challenges, our results underscore the gap between self-supervised and centralized methods and suggests improving self-supervised MARL methods as a meaningful future direction.
    
\label{disc}

\clearpage
\bibliographystyle{unsrtnat}
\bibliography{ref}  


\clearpage
\section{Supplementary Material}
\label{supp}
\paragraph{Environmental steps during training} All models are trained up to one million environment steps by dividing them into 200 epochs, each with 5000 steps in the environment. To compare across models, we select the best-performing epoch on the test episodes, and list the number of steps occurre in Table \ref{tab:cleanup} for Clean Up and Table \ref{tab:harvest} for Harvest.
\begin{table}[h]
\centering
\caption{Training Steps - Clean UP}
\label{tab:cleanup}
\begin{tabular}{|l|c|}
\hline
Model & Environment Steps \\ \hline
IPPO   &    762000           \\ \hline
ICM-R     &    276000             \\ \hline
ICM    &  348000            \\ \hline
Influence   &   726000            \\ \hline
SVO-HE    &     222000         \\ \hline
SVO-HO    &     258000         \\ \hline
MAPPO    &     255000         \\ \hline
\end{tabular}
\end{table}

\begin{table}[h]
\centering
\caption{Training Steps - Harvest}
\label{tab:harvest}
\begin{tabular}{|l|c|}
\hline
Model & Environment Steps \\ \hline
IPPO   &    594000           \\ \hline
ICM-R    &   1000000           \\ \hline
ICM    &    816000          \\ \hline
Influence   &     966000          \\ \hline
SVO-HE    &    900000          \\ \hline
SVO-HO    &    690000          \\ \hline
SVO-HO    &    630000          \\ \hline
\end{tabular}
\end{table}

Implementation for this work can be found at \url{https://github.com/violetxi/EmergentSocialDynamics}


\end{document}